\documentclass[superscriptaddress,pra,twocolumn,amsmath,amssymb,showkeys,showpacs]{revtex4-1}

\usepackage{graphicx}
\usepackage{graphics}
\usepackage{amssymb}
\usepackage{amsmath,bm}
\usepackage{float}
\usepackage[usenames,dvipsnames]{color}
\usepackage[export]{adjustbox}
\usepackage{hyperref}
\hypersetup{
    colorlinks=true,
    linkcolor=blue,
    citecolor=blue,
    filecolor=cyan,      
    urlcolor=cyan,
}
\begin{document}

\title{Spatial rogue waves in photorefractive SBN crystals}

\author{C. Hermann-Avigliano}\email{Corresponding author email: carla.hermann@uchile.cl}
\affiliation{Departamento de F\'{i}sica and Millennium Institute for Research in Optics (MIRO), Facultad de Ciencias and Facultad de Ciencias F\'{i}sicas y Matem\'{a}ticas, Universidad de Chile, Santiago, Chile}

\author{I.A. Salinas}
\affiliation{Departamento de F\'{i}sica and Millennium Institute for Research in Optics (MIRO), Facultad de Ciencias and Facultad de Ciencias F\'{i}sicas y Matem\'{a}ticas, Universidad de Chile, Santiago, Chile}

\author{D.A. Rivas}
\affiliation{Departamento de F\'{i}sica and Millennium Institute for Research in Optics (MIRO), Facultad de Ciencias and Facultad de Ciencias F\'{i}sicas y Matem\'{a}ticas, Universidad de Chile, Santiago, Chile}

\author{B. Real}
\affiliation{Departamento de F\'{i}sica and Millennium Institute for Research in Optics (MIRO), Facultad de Ciencias and Facultad de Ciencias F\'{i}sicas y Matem\'{a}ticas, Universidad de Chile, Santiago, Chile}

\author{A. Man\v{c}i\'c}
\affiliation{Dept. of Phys., Faculty of Sciences and Math., University of Ni\v s, Serbia}

\author{C. Mej\'ia-Cort\'es}
\affiliation{Programa de F\'{i}sica, Facultad de Ciencias B\'{a}sicas, Universidad del Atl\'{a}ntico, Barranquilla, Colombia}

\author{A. Maluckov}
\affiliation{Vinca Institute of Nuclear Sciences, University of Belgrade, Serbia}

\author{R.A. Vicencio}
\affiliation{Departamento de F\'{i}sica and Millennium Institute for Research in Optics (MIRO), Facultad de Ciencias and Facultad de Ciencias F\'{i}sicas y Matem\'{a}ticas, Universidad de Chile, Santiago, Chile}

\begin{abstract}
We report on the excitation of large-amplitude waves, with a probability of around $1\%$ of total peaks, on a photorefractive SBN crystal, by using a simple experimental setup at room temperature. We excite the system using a narrow gaussian beam and observe different dynamical regimes tailored by the value and time rate of an applied voltage. We identify two main dynamical regimes: a caustic one for energy spreading and a speckling one for peak emergence. Our observations are well described by a two-dimensional Schr\"odinger model with saturable local nonlinearity. \url{https://www.osapublishing.org/ol/abstract.cfm?uri=ol-44-11-2807}
\end{abstract}

\maketitle

The phenomenon of rogue waves (RWs) dates back from observations of isolated large amplitude water waves on the sea surface, appearing out of nowhere and disappearing without a trace~\cite{ocean}. These rare events were statistically associated with long tails of high amplitude wave distributions. Nowadays, they are related to extreme events (EEs) arising in the presence of many uncorrelated grains of activity, which are inhomogeneously distributed in large spatial domains of complex media and they are being studied in different fields of science~\cite{general}. Diversity and particularity of RWs cause many uncertainties regarding their definition, origin, predictability and statistics~\cite{review}. Besides the fact that EEs generation is related to the phenomena with long-tails statistics, there is a well-established approach that relates their appearance with the merging of coherent structures~\cite{coherent1,coherent2}. Within optical systems, the study of RWs include light propagation in optical fibers~\cite{coherent1}, nonlinear optical cavities~\cite{cavity}, and photorefractive crystals. RWs were observed on a BaTiO$_3$:Co crystal~\cite{crystals1} on a highly nonlinear regime showing spatiotemporal turbulence. Ref.~\cite{crystals2} reports optical RWs on a KLTN crystal where nanodisorder, giant nonlinearity (NL) and high temperature generate large intensity events (IEs). The ferroelectric-to-paraelectric transition also generate RWs due to thermally induced focusing and defocusing effects~\cite{PRLSBN}, where a transition from linear to highly nonlinear regimes promotes a turbulent dynamics~\cite{pier2016}. Modulational instability (MI) and pattern formation in SBN photorefractive crystals, using incoherent~\cite{moti1} and coherent~\cite{jeng1} light beams, promote the appearance of different optical patterns, including stripes and filaments. This is due to a combined action of different mechanisms like crosstalk and NL, which are believed to be essential for the observation of RWs~\cite{crystals1}. It was shown recently that the existence of purely linear large IEs was also possible due to isolated caustic effects on an optical sea~\cite{caustic}.

Here, we investigate experimentally, and corroborate numerically, the appearance of large-amplitude events on a SBN photorefractive crystal. The key elements in our study are the simplicity of the experiment, its reproducibility, and the robust appearance of RWs under simple controlled conditions, without requiring large NLs neither turbulence phenomena. By injecting a low power Gaussian beam (GB), and ramping an externally applied voltage, we are able to distinguish between different dynamical regimes. We observe that for low applied voltages (weak NL) the beam experiences a caustic-like distribution. Amplitudes at the background level are very small and, therefore, we observe a mixture of linear and nonlinear waves coexisting and forming different interference profiles, which resemble caustic-like patterns~\cite{caustic}. When increasing the voltage and, as a consequence the NL of the system, we observe a pattern fragmentation into narrow light spots, where some of them have a huge intensity. We compare the peak intensities measured in the experiment and in the numerical simulations to identify the excitation of EEs. Additionally, we numerically study the dynamics along the crystal to elucidate the appearance of RWs during propagation.

\begin{figure}[htbp]
\centering
\includegraphics[height=4cm, width=0.4\textwidth]{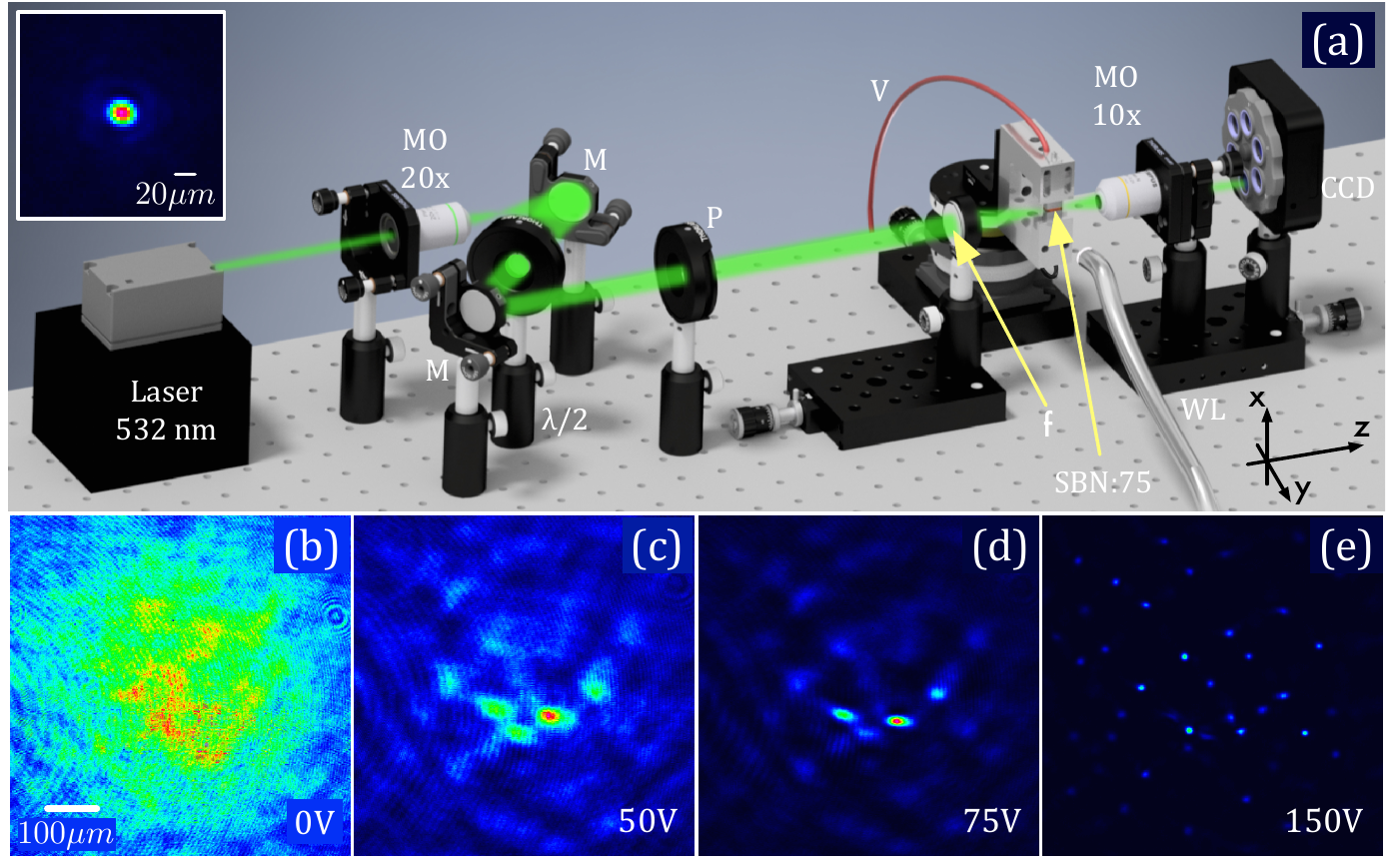}
\caption{(a) Experimental setup for observing RWs. Inset: Input profile. (b)--(e) Output beam profiles for indicated voltages. In (b) $V=0$.}
\label{fig1}
\end{figure}

We start our investigation by experimentally studying the propagation of GBs on a $0.005\%$ CeO$_2$ doped SBN:75 photorefractive crystal, using the setup sketched in Fig.~\ref{fig1}(a). Our sample has a transversal area of $5\times2$ mm$^2$ with a length of $10$ mm (propagation coordinate $z$). Our crystal has a $20$ times larger nonlinear response for a polarization along the vertical axis ($x$), in comparison to the polarizations perpendicular to it~\cite{sbn}. The nonlinear response is controlled indirectly by an external voltage applied in the crystal vertical axis (smaller crystal dimension). We expand a $532$ nm laser beam by a $20\times$ microscope objective. We define the beam polarization ($x$) and power (10 $\mu$W) with a sequence of a $\lambda/2$ waveplate and a linear polarizer. The optical axis is defined by aligning the laser beam with respect to the sample center. The crystal is fixed to a rotational-elevation stage mounted on a micrometer translation platform. To observe the optical patterns at the input and output facets of the crystal, we use a beam profiler CCD camera.

We explore the crystal response by defining two input parameters: beam width and power. The former is reduced by using a plano convex lens of $50$ mm focal length, achieving an input GB approximately $10\ \mu$m wide (see Fig.~\ref{fig1}(a)-inset). The output beam profile after propagating $10$ mm through the crystal without any applied voltage, is shown in Fig.~\ref{fig1}(b). As it is observed, the input beam strongly spreads inside the crystal and it spatially evolves to a wide gaussian profile, $\approx50$ times larger. This pattern possesses several spatial irregularities, which can be related either to fluctuations in the nominal linear refractive index (due to previous experiments) or to inhomogeneities on the GB itself. These spatial fluctuations are important in our experiment because they create spatial regions with different light density and, in someway, they initialize the filamentation process. Therefore, we have a natural spatial symmetry breaking mechanism, which drives the system to a non-homogeneous state.

Our experiment shows rich dynamics depending on the parameters used to excite the SBN sample. This is related to the nonlinear response of the crystal, which in our case depends directly on the light intensity. However, as we completely define the light intensity to a constant value (fixing the input power and input waist), we effectively modify the nonlinear response of the crystal by means of an externally applied voltage. When increasing this control parameter, we observe two different dynamical regimes. If we increase the voltage fast enough ($\sim$ seconds), the GB rapidly collapses to a 2D bright soliton~\cite{2dsoliton}. After this solution is formed, it is possible to observe stable or unstable patterns, which strongly depend on the input power, external voltage and crystal length. Differently, a slow voltage increment ($\sim$ minutes) allows for the light to spread smoothly over the crystal, facilitating the creation of several regions of larger light density, observing some kind of agglomeration diffusive process. Interestingly, here light is able to localize weakly over different spatial domains. Nevertheless, as the local power is not that high, the radiation of energy between neighboring big spots (crosstalk) is still possible. This smooth dissemination of energy (mediated by an inhomogeneous initial diffraction process), plus a slow nonlinear increment, gives us the necessary mechanisms to observe large-amplitude events on our photorefractive setup, without requiring to work neither on a highly nonlinear regime~\cite{crystals1} nor close to a thermal crystal transition~\cite{crystals2}.

Typical output profiles for the slow variation regime are shown in Figs.~\ref{fig1}(c)--(e). To statistically analyze the data, we defined the following protocol: we increase the voltage from $0$ to a maximum of $225$ V, in steps of $25$ V every $15$ minutes. This smooth increment allows an adiabatic transformation of the output spatial profile. We observe a rather static (nonlinear stationary-like complex localized) pattern that allows us to take a representative image of the output facet every $15$ minutes, which characterizes the state of the system at a given voltage. We obtain images every $25$ V, although we focus on larger voltage values where peaks are spatially more localized, having an average width lower than $\approx 10\ \mu$m. At low voltages, we observe a tendency of a macroscopic agglomeration of energy in wide light spots, as Figs.~\ref{fig1}(c) and (d) show. Then, by a further increment of the voltage up to $\sim 100$ V, we observe that narrow light spots become connected by some kind of \textit{light currents} that are associated to a caustic-like energy spreading~\cite{caustic}. By increasing further the voltage, we observe the appearance of several large amplitude peaks, which have a small individual spatial extent [see Fig.~\ref{fig1}(e)]. We run the same experiment $30$ times to increase the statistical ensemble. Before initializing every new experimental realization, we apply a white light source to erase any induced refractive index pattern, which could be imprinted inside the crystal in a previous experiment. We check this by inspecting the output linear profile [Fig.~\ref{fig1}(b)] and determine whether there is a need to continue erasing the crystal or to simply translate the sample to a more homogeneous region.

\begin{figure}[htbp]
\centering
\includegraphics[height=4cm, width=0.45\textwidth]{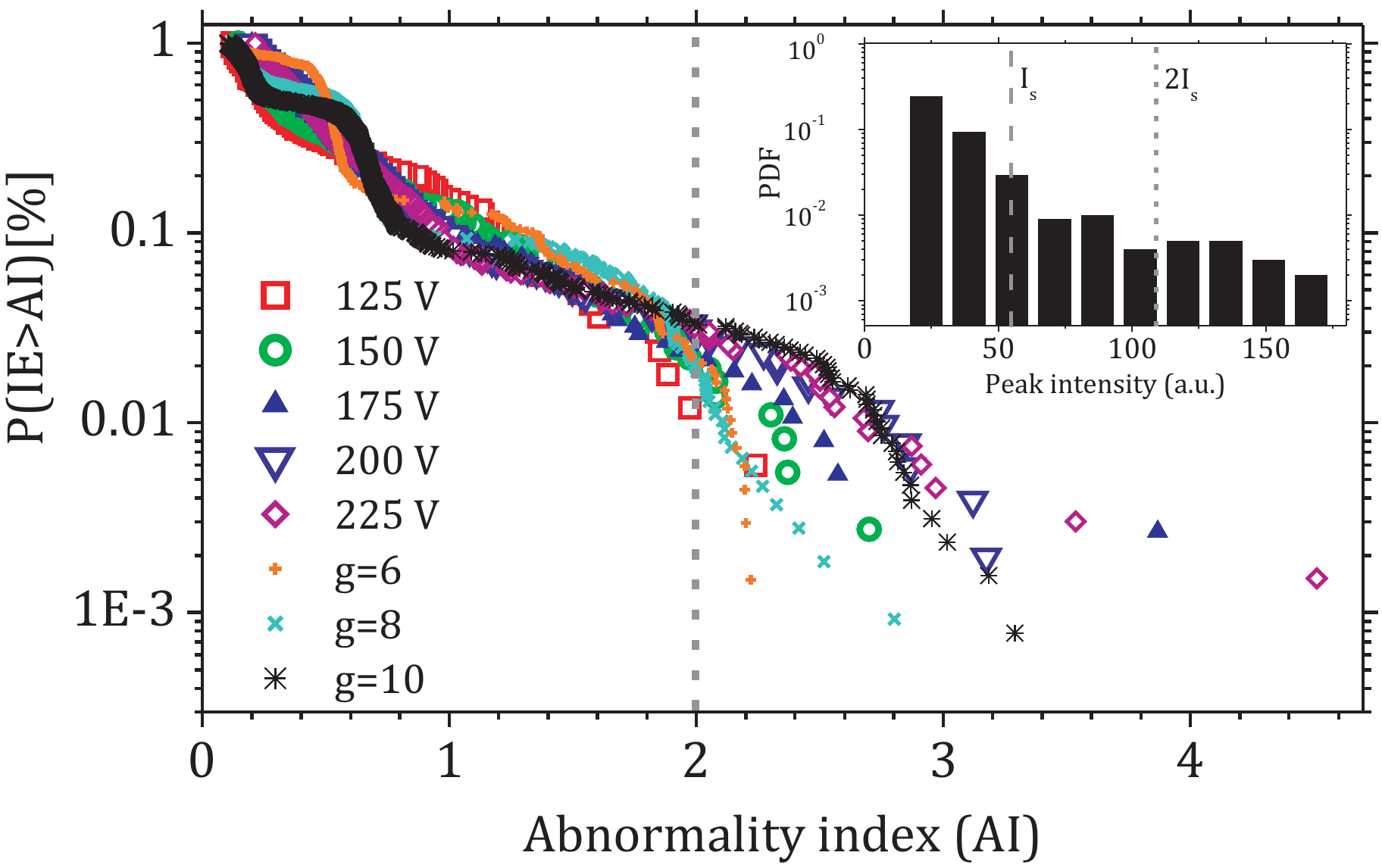}
\caption{Semi-log plot of the probabilities (in $\%$) of having an IE above a particular $AI$ for different voltages. All events above $AI=2$ are consider as EEs (vertical dotted line). The numerical counterpart for $g=6$, $8$ and $10$ is presented as well. Inset: PDF for $200\,V$. $I_s$ is represented by the vertical dashed line.}
\label{fig2}
\end{figure}
%

We analyze each obtained image and look for local maxima above a given defined threshold value (chosen to avoid background events). To avoid saturation, we set the exposure time on the beam profiler to $1$ ms for $125$ and $150$ V, and to $0.5$ ms for range $175-225$ V. The intensity scale is defined in the interval $0-255$ levels (typical scale for images), where zero means no light and $255$ represents the largest intensity, depending on the chosen exposure time. In general,  heavy-tailed intensity distributions are an indicator of the existence of EEs~\cite{26}. By following a standard criterion on RWs~\cite{coherent2}, we consider as EEs those with intensities larger than twice a significant intensity $I_s$, which is defined as the average value of the highest intensity tertile of the corresponding probability density function (PDF) distribution (see inset in Fig.~\ref{fig2}). Events with an abnormality index $AI\equiv I/I_s>2$ are considered as RWs. To determine the probability of RWs, we compute $P($IE$>AI)=1-$ cumulative PDF.  This represent the probability of having an IE with an $AI$ larger than a certain value, $P($IE$>AI)$ (see Fig.~\ref{fig2}).  Hence, the probability of having RWs corresponds to the value at which the data crosses $AI=2$. We detect for $\{125,150,175,200,225\}$  voltages, a total number of $\{167,364,375,511,662\}$ IE, from where $\{1,7,8,16,20\}$ are considered as EEs, respectively. Therefore, the percentage of occurrence in our experiments is only $\{0.6,1.9,2.1,3.1,3.0\}\%$. We observe that large intensity events are always below $\sim3\%$ of the total data, which indicates that our reported RWs are rare and have very low statistics. 

From the theoretical side, the light propagation through photorefractive media can be modeled mathematically by a $2$D nonlinear Schr\"{o}dinger equation with saturable nonlinearity~\cite{sbn}
%
\begin{equation}
i\frac{\partial}{\partial z} \psi(x,y,z)+\beta
\nabla_{\bot}\psi(x,y,z)-g\frac{\psi(x,y,z)}{1+|\psi(x,y,z)|^2}=0\ .
\label{eq1}
\end{equation}
%
$\psi(x,y,z)$ corresponds to the envelope of the electrical field, $z$ to the propagation coordinate, $\beta$ the diffraction coefficient (fixed to $1$), and $\nabla_{\bot}$ corresponds to the transverse Laplacian operator. The nonlinear coefficient is denoted by $g$ and it is proportional to the external applied voltage in the experiment. A positive $g$-value implies a focusing regime, while a negative one refers to a defocusing case~\cite{review}. We focus here on describing phenomenologically the observed steady-state patterns, which are in general well described by model~(\ref{eq1}) (local saturable NL is able to produce long-range phenomena, due to a natural generation of broader spatial patterns~\cite{sbn}). The quantitative changes between the two regimes can be associated with the slow nonlinear response of the photorefractive SBN crystal, which is of saturable nature. This response is determined by the interplay between external voltage, charge dynamics and light interaction. Under the influence of internal and external electric fields, charges start to redistribute, hence a non-homogeneous refractive index change occurs via the electro-optic effect. Then, the different observed dynamical phases could be related to different values and rates of the external voltage in the experiment.

\begin{figure}[t!]
\centering
\includegraphics[height=4cm, width=0.45\textwidth]{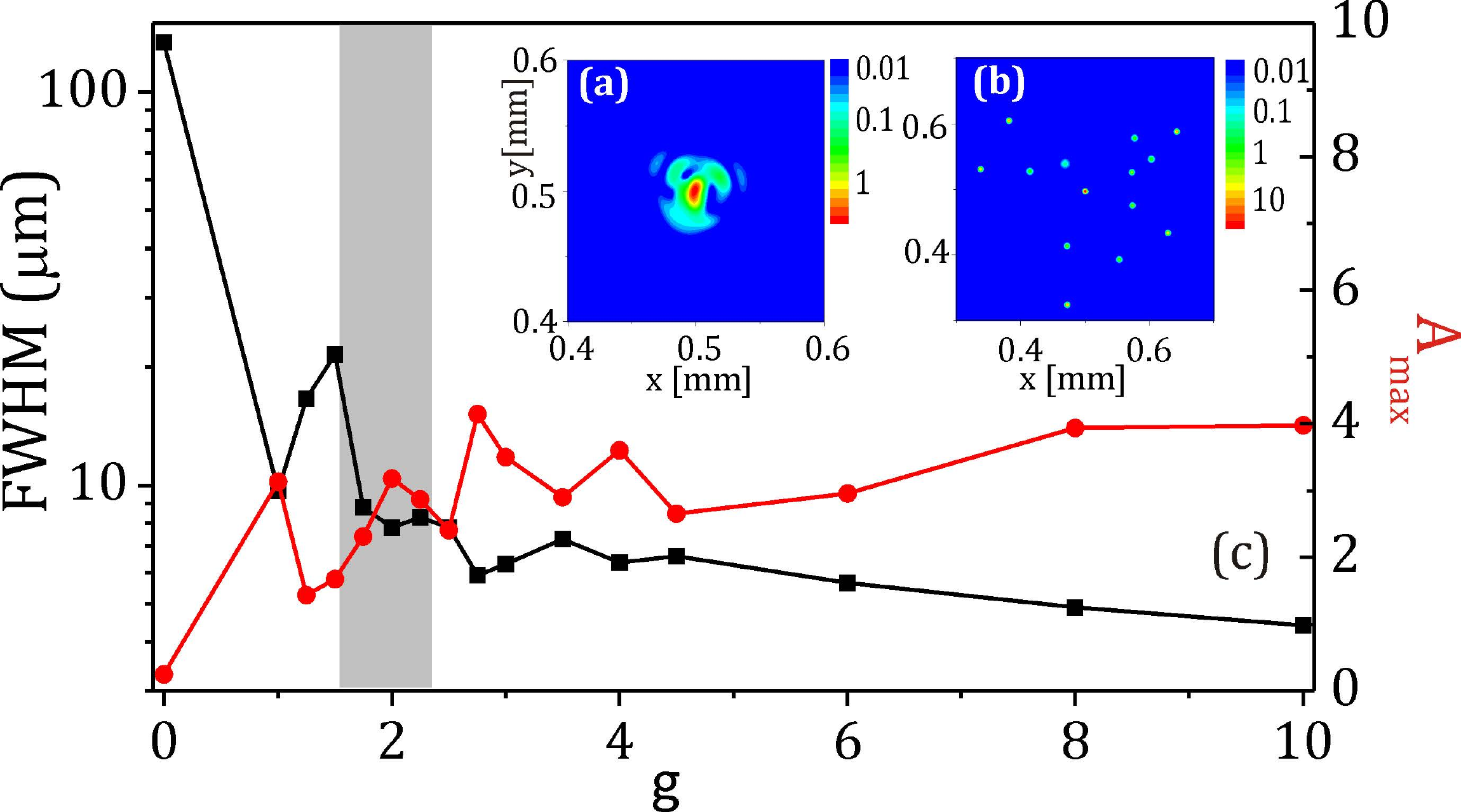}
\caption{Numerically output beam intensity profiles for $g=1.75$ (a) and $g=8$ (b). (c) Average FWHM (log-linear scale) and $A_{max}$ vs $g$, for the output profile $\psi(x,y,z_{end})$. Shaded area indicates a caustic-like regime region.}
\label{fig3}
\end{figure}

We initialize our numerical simulations by injecting a noise seeded GB, with a certain width and intensity (similar to experimental values), into equation~(\ref{eq1}). The noise is added by an uniform random number generator, having a mean value equal to zero and a maximum value equal to $1\%$ of the beam amplitude. The beam propagation through the SBN saturable media is obtained by applying a standard split-step pseudo-spectral  procedure~\cite{nash}. We fix the propagation length to $10$ mm, while the input width is of the order of $10$ $\mu$m. The transversal area is $1\times1$ mm$^2$. Different dynamical regimes can be distinguished in the parameter region above the MI threshold, depending on $g$ (Fig. \ref{fig3}). Randomly fluctuating intensity patterns are obtained for low $g$-values (Fig.~\ref{fig3}(a)), which agrees well with the experiments at low voltage (Fig.~\ref{fig1}(c)). We identify a wide agglomeration, having a non-homogeneous pattern due to the effects associated to NL during the beam propagation. When $g$ is increased, the beam shows a speckling-like profile due to the initial non-homogeneous expansion of the light and the corresponding self-focusing response of the crystal. This allows the clustering of light in very specific regions generating high intensity spots. Figure~\ref{fig3}(b) shows a numerical example of this for $g=8$, which is in good phenomenological agreement with the experimental results shown in Fig.~\ref{fig1}(e). 

The numerical integration provides the possibility to construct a phase diagram to study different dynamical regimes, with $g$ as a control parameter. Figure~\ref{fig3}(c) shows the averaged full width half maximum (FWHM) of output profiles versus $g$. We observe in a purely linear regime that the FWHM expands strongly compared to the input value. In the presence of NL, the beam simply shrinks due the self-trapping phenomena and the output FWHM just decreases. However, a transient regime between $1\lesssim g\lesssim1.7$ can be distinguished. Here, a single peak soliton-like beam generates wide multi-peak localized structures. This precedes the regime with caustic-like patterns ($1.7\lesssim g\lesssim2.5$), characterized by an averaged FWHM of around $7.5$ $\mu$m (shaded area in Fig.~\ref{fig3}(c)). After this caustic region, the averaged FWHM is reduced, which is consistent with the appearance of small size light spots due to the overall filamentation process occurring at larger values of $g$. We simultaneously plot the maximum amplitude ($A_{max}\equiv\sqrt{I_{max}}$) observed for every realization, after averaging at different values of $g$. We observe that this amplitude has a slight increment when the dynamics change from a caustic regime to a filamentation process. This is due to conservation of energy, which indicates that a multi-peak pattern has to split the total power in several spots and, therefore, peaks can only contain a limited amount of energy. Additionally, the saturable nature of the NL produces an upper bound for $A_{max}$. 

By statistically analyzing the output intensity profiles, we compare our numerical simulations with the data in Fig.~\ref{fig2}.  We check the distribution of high intensity peaks for $100$ numerical realizations for $g=6$, $8$ and $10$ (speckling regime), and compute as well the $P($IE$>AI)$. These results are presented as additional confirmation of suitability of a relatively simple mathematical-numerical model to our experiment, with the main objective of demonstrating a qualitative comparison of the phenomena. For $g$ equal to $\{6,8,10\}$ we find $\{14,20,42\}$ EEs on an ensemble of $\{676,1080,1281\}$ IE. We observe that RWs increase with NL, as expected from the experimental counterpart. 

\begin{figure}[t!]
\centering
\includegraphics[height=4cm, width=0.45\textwidth]{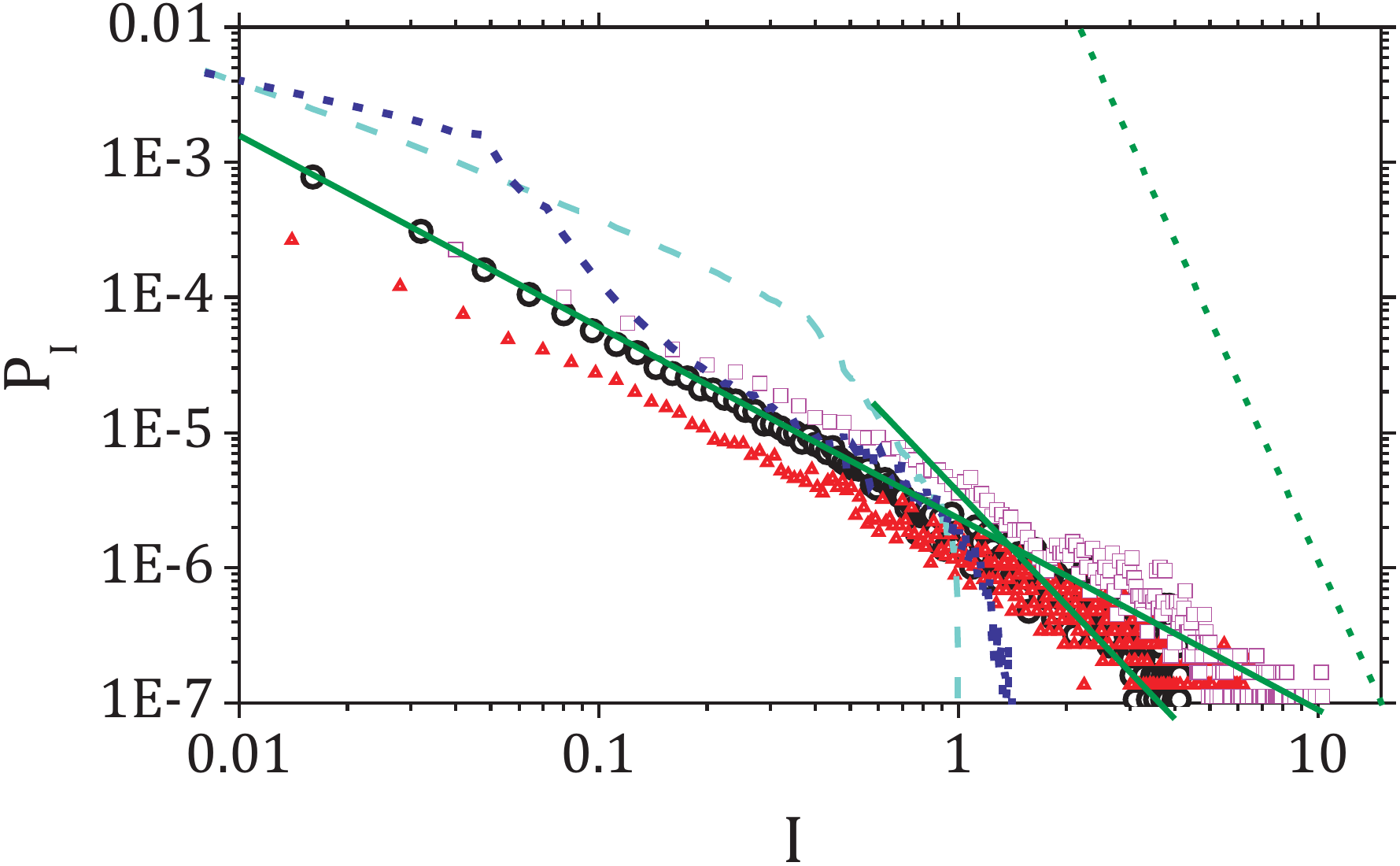}
\caption{Log-Log plot of $P_I$ vs $I$, for $g = 0.1,\, 0.5$ (no RWs), $1.85$ (caustic-like regime), $6$ and $10$ (speckles regime) represented by light blue, purple dashed lines, black circles, red triangles, and magenta empty. Dark green dashed line corresponds to a $-1$ slope.} 
\label{fig4}
\end{figure}

The experimental data in our setup is collected from the output intensity profiles, hence the dynamics inside the crystal is  "hidden". However, numerical simulations allow us to study the dynamics along $z$. We compute the integral intensity distribution $P_I$~\cite{ranije} for different values of $g$, which takes into account light intensities along the whole propagation through the sample (Fig.~\ref{fig4}). For $g=0.1$ and $0.5$ the tails are insignificant indicating the negligible number of high amplitude events. The slope of tails is exponential one ($<-1$). Entering into the caustic-like region, $g = 1.85$, we observe a power-law behavior of $P_I$ in the whole range of intensities. We observe that the slope of this distribution changes with $I$ (see the two dark green straight lines in Fig.~\ref{fig4}). Power-law like distributions are observed in the filamentation (speckling) regime too, with different slope rates ($g=\{6,10\}$), commonly related to the emergence of EEs~\cite{review}. Differently, the maximum intensities in the caustic-like regime are smaller compared to those reached in the speckling regime for the chosen $g$ values. For $g = 10$ we observe that there are more events with smaller intensity than for $g = 1.85$, as a consequence of slow and spontaneously initiated clustering of events in the background. These clusters represent energy seeds for the formation of few high IEs. Additionally, we can estimate the probability of occurrence of EEs, defined as $Pee = \int_{ I>Ia} P_I dI$~\cite{ranije}, obtaining that $P_{ee}$  is of order $0.1\%$ to $1\%$ in caustic-like and speckling cases presented here.
This value agrees well with the experimentally and numerically estimated ratio (EE/IE) extracted from maximum light
intensity at the output facet of the crystal. The last points out the existence and significance of huge amplitude events during the whole light propagation through the crystal. We plan to continue our research in this direction expecting to arrange a new experimental setup to confirm our numerical findings.

We have studied experimental and numerically the appearance of large amplitude waves on a SBN crystal. By tuning the external voltage, we were able to generate the experimental conditions to observe different dynamical regimes. We identified a caustic-like regime, where an effective energy dissemination through the crystal was observed. When increasing the voltage, we observed the growth of local clusters with the consequent generation of large amplitude peaks having very low statistics. This is always observed in our setup, showing that large IEs are quite a robust phenomena in these kind of nonlinear systems and, particularly in our experiment, without the need of any extra mechanism neither any special stimulation. The key elements in our study are the simplicity of the experiment, its reproducibility, and the robust appearance of RWs under simple conditions. Our observations are well supported by a saturable model, which correctly predicts the dissemination and the speckling regimes. Our numerics give also an estimation of the occurrence of RWs in the order of $1\%$ of total peaks, which agrees well with our experimental results. 

\textbf{Funding.} Fondo Nacional de Desarrollo Cient\'ifico y Tecnol\'ogico (FONDECYT) (1151444, 3180153); Programa ICM Millennium Institute for Research in Optics (MIRO); U-Inicia VID Universidad de Chile (UI 004/2018); Ministarstvo Prosvete, Nauke i Tehnolo\v{s}kog Razvoja, Republike Srbije (III 45010); National laboratory for high performance computing (ECM-02), University of Chile.

\textbf{Acknowledgment.} Authors acknowledge C. Cantillano, M.G. Clerc and G. Gonz\'alez for useful discussions. \\

\textbf{© [2019] Optical Society of America. One print or electronic copy may be made for personal use only. Systematic reproduction and distribution, duplication of any material in this paper for a fee or for commercial purposes, or modifications of the content of this paper are prohibited.}


\begin{thebibliography}{0}%
\makeatletter
\providecommand \@ifxundefined [1]{%
 \@ifx{#1\undefined}
}%
\providecommand \@ifnum [1]{%
 \ifnum #1\expandafter \@firstoftwo
 \else \expandafter \@secondoftwo
 \fi
}%
\providecommand \@ifx [1]{%
 \ifx #1\expandafter \@firstoftwo
 \else \expandafter \@secondoftwo
 \fi
}%
\providecommand \natexlab [1]{#1}%
\providecommand \enquote  [1]{``#1''}%
\providecommand \bibnamefont  [1]{#1}%
\providecommand \bibfnamefont [1]{#1}%
\providecommand \citenamefont [1]{#1}%
\providecommand \href@noop [0]{\@secondoftwo}%
\providecommand \href [0]{\begingroup \@sanitize@url \@href}%
\providecommand \@href[1]{\@@startlink{#1}\@@href}%
\providecommand \@@href[1]{\endgroup#1\@@endlink}%
\providecommand \@sanitize@url [0]{\catcode `\\12\catcode `\$12\catcode
  `\&12\catcode `\#12\catcode `\^12\catcode `\_12\catcode `\%12\relax}%
\providecommand \@@startlink[1]{}%
\providecommand \@@endlink[0]{}%
\providecommand \url  [0]{\begingroup\@sanitize@url \@url }%
\providecommand \@url [1]{\endgroup\@href {#1}{\urlprefix }}%
\providecommand \urlprefix  [0]{URL }%
\providecommand \Eprint [0]{\href }%
\providecommand \doibase [0]{http://dx.doi.org/}%
\providecommand \selectlanguage [0]{\@gobble}%
\providecommand \bibinfo  [0]{\@secondoftwo}%
\providecommand \bibfield  [0]{\@secondoftwo}%
\providecommand \translation [1]{[#1]}%
\providecommand \BibitemOpen [0]{}%
\providecommand \bibitemStop [0]{}%
\providecommand \bibitemNoStop [0]{.\EOS\space}%
\providecommand \EOS [0]{\spacefactor3000\relax}%
\providecommand \BibitemShut  [1]{\csname bibitem#1\endcsname}%
\let\auto@bib@innerbib\@empty
\end{thebibliography}%


\begin{thebibliography}{99}

\bibitem{ocean} C. Kharif and E. Pelinovsky, ``Physical mechanisms of the rogue wave phenomenon,'' Eur. J. Mech. B/Fluids \textbf{22}, 603 (2003).
\bibitem{general} S. Albevario, V. Jentsch, and H. Kantz, ``Extreme Events in Nature and Society", Springer, Berlin, 2006.
\bibitem{review} M. Onorato, S. Residori, U. Bortolozzo, A. Montina, and F.T. Arecchi,``Rogue waves and their generating mechanisms in different physical context,'' Phys. Rep. \textbf{528}, 47 (2013).
\bibitem{coherent1} D. R. Solli, C. Ropers, P. Koonath and B. Jalali, "Optical rogue waves", Nature 450, 1054 (2007).
\bibitem{coherent2} Marcel G. Clerc, Gregorio Gonz\'{a}lez-Cort\'{e}s, and Mario Wilson, "Extreme events induced by spatiotemporal chaos in experimental optical patterns", Opt. Lett. \textbf{41}, 2711-2714 (2016), and references within.
\bibitem{cavity} A. Montina, U. Bortolozzo, S. Residori, and F. T. Arecchi, "Non-Gaussian Statistics and Extreme Waves in a Nonlinear Optical Cavity", Phys. Rev. Lett. \textbf{103}, 173901 (2009).
\bibitem{crystals1} N. Marsal, V. Caullet, D. Wolfersberger, and M. Sciamanna, "Spatial rogue waves in a photorefractive pattern-forming system," Opt. Lett. \textbf{39}, 3690-3693 (2014).
\bibitem{crystals2} D. Pierangeli, F. Di Mei, C. Conti, A. J. Agranat, and E. DelRe, "Spatial Rogue Waves in Photorefractive Ferroelectrics", Phys. Rev. Lett. \textbf{115}, 093901 (2015).
\bibitem{PRLSBN}M. O. Ram\'irez, D. Jaque, L. E. Baus\'a, J. Garc\'ia Sol\'e, and A. A. Kaminskii, ``Coherent Light Generation from a Nd:SBN Nonlinear Laser Crystal through its Ferroelectric Phase Transition,'' Phys. Rev. Lett. \textbf{95}, 267401 (2005).
\bibitem{pier2016}D. Pierangeli,F. Di Mei, G. Di Domenico, A.J. Agranat, C. Conti, and E. DelRe,``Turbulent Transitions in Optical Wave Propagation,'' Phys. Rev. Lett. \textbf{117}, 183902 (2016).
\bibitem{moti1}D. Kip, M. Soljacic, M. Segev, E. Eugenierva, and D.N. Christodoulides, ``Modulational instability and pattern formation in spatially incoherent light beams,'' Science \textbf{290}, 495 (2000).
\bibitem{jeng1}C.-C. Jeng, Y. T. Lin, R.-C. Hong, and R.-K. Lee, ``Optical pattern transitions from modulation to transverse instabilities in photorefractive crystals,'' Phys. Rev. Lett. \textbf{102}, 153905 (2009).
\bibitem{caustic}A. Mathis, L. Froehly, S. Toenger, F. Dias, G. Genty and John M. Dudley, ``Caustic and rogue waves in an optical sea,'' Sci. Rep. \textbf{5}, 12822 (2015).
\bibitem{sbn}R. Allio, D. Guzm\'an-Silva, C. Cantillano, L. Morales-Inostroza, D. L\'opez-Gonz\'alez, S. Etcheverry, R. A. Vicencio, and J. Armijo, "Photorefractive writing and probing of anisotropic linear and nonlinear lattices," J. Opt. \textbf{17}, 025101 (2015).
\bibitem{2dsoliton}S. Lan, M. F. Shih and M. Segev, "Self-trapping of one-dimensional and two-dimensional optical beams and induced waveguides in photorefractive KNbO(3)", Opt. Lett. \textbf{22}, 1467 (1997).
\bibitem{26} E. Louvergneaux, V. Odent, M.I. Kobolov, and M. Taki,"Statistical Analysis of spatial frequency supercontinuum in pattern forming feedback system", Phys.Rev. A \textbf{87}, 063802 (2013).
\bibitem{nash} F.~Chen, M.~Stepi\'c, C. E.~R\"uter, D.~Runde, D.~Kip, V.~Shandarov, O.~Manela, and M.~Segev, "Discrete diffraction and spatial gap solitons in photovoltaic LiNbO3 waveguide arrays,"  Opt. Express \textbf{426} 13, 4314 (2005).
\bibitem{ranije} A. Maluckov, Lj. Had\v zievski, N. Lazarides, and G. P. Tsironis, "Extreme events in discrete nonlinear lattices," Phys. Rev. E \textbf{79}, 025601R (2009). 


\end{thebibliography}
\end{document}